\documentclass{aastex631}

\usepackage{rotating}

\newcommand{\Msun}{\mbox{$\mathrm{M}_{\odot}$}}

\newcommand{\kms}{km\,s$^{-1}$}
\newcommand{\Teff}{\mbox{$T_{\mathrm{eff}}$}}
\newcommand{\logg}{\mbox{$\log g$}}
\newcommand\lya{$\mathrm{Lyman}\,\alpha$}
\newcommand\eguma{EG\,UMa}

\shorttitle{The Lyman $\alpha$ line of EG\,UMa}
\shortauthors{Wilson et al.}
\graphicspath{{./}{figures/}}
\begin{document}

\title{Testing \lya\ emission line reconstruction routines at multiple velocities in one system.}

\correspondingauthor{David J. Wilson}
\email{djwilson394@gmail.com}

\author[0000-0001-9667-9449]{David J. Wilson}
\affil{Laboratory for Atmospheric and Space Physics, University of Colorado at Boulder, 600 UCB, Boulder, CO 80303}
% \affil{McDonald Observatory, University of Texas at Austin, Austin, TX 78712}

\author[0000-0002-1176-3391]{Allison Youngblood}
%\altaffiliation{NASA Postdoctoral Program Fellow}
\affil{Goddard Space Flight Center, Greenbelt, MD 20771}
\affil{Laboratory for Atmospheric and Space Physics, University of Colorado at Boulder, 600 UCB, Boulder, CO 80303}

\author[0000-0002-2398-719X]{Odette Toloza}
\affil{Departamento de Física, Universidad Técnica Federico Santa María, Avenida España 1680, Valparaíso, Chile}

\author[0000-0002-0210-2276]{Jeremy J.  Drake}
\affil{Center for Astrophysics | Harvard \& Smithsonian, 60 Garden Street, Cambridge, MA 02138, USA}

\author[0000-0002-1002-3674]{Kevin France}
\affil{Laboratory for Atmospheric and Space Physics, University of Colorado at Boulder, 600 UCB, Boulder, CO 80303}

\author[0000-0001-8499-2892]{Cynthia S. Froning}
\affil{McDonald Observatory, University of Texas at Austin, Austin, TX 78712}

\author[0000-0002-2761-3005]{Boris T. G{\"a}nsicke}
\affil{Department of Physics, University of Warwick, Coventry CV4 7AL, UK}

\author[0000-0003-3786-3486]{Seth Redfield}
\affil{Wesleyan University, Department of Astronomy and Van Vleck Observatory, 96 Foss Hill Dr., Middletown, CT 06459, USA}

\author[0000-0002-4998-0893]{Brian E. Wood}
\affil{Naval Research Laboratory, Space Science Division, Washington, DC 20375, USA}



\begin{abstract}
The 1215.67\,\AA\ \ion{H}{1} \lya\ emission line dominates the ultraviolet flux of low mass stars, including the majority of known exoplanet hosts. Unfortunately, strong attenuation by the interstellar medium (ISM) obscures the line core at most stars, requiring the intrinsic \lya\ flux to be reconstructed based on fits to the line wings. We present a test of the widely-used \lya\ emission line reconstruction code {\sc lyapy} using phase-resolved, medium-resolution STIS G140M observations of the close white dwarf-M dwarf binary \eguma. The Doppler shifts induced by the binary orbital motion move the \lya\ emission line in and out of the region of strong ISM attenuation. Reconstructions to each spectrum should produce the same \lya\ profile regardless of phase, under the well-justified assumption that there is no intrinsic line variability between observations. Instead, we find that the reconstructions underestimate the \lya\ flux by almost a factor of two for the lowest-velocity, most attenuated spectrum, due to a degeneracy between the intrinsic \lya\ and ISM profiles. Our results imply that many stellar \lya\ fluxes derived from G140M spectra reported in the literature may be underestimated, with potential consequences for, for example, estimates of extreme-ultraviolet stellar spectra and ultraviolet inputs into simulations of exoplanet atmospheres. 
\end{abstract}

% \keywords{}


\section{Introduction} \label{sec:intro}
The \ion{H}{1} \lya\ line at 1215.67\,\AA\ is a paradox of stellar observational astronomy: Vitally important for the study of stellar atmospheres and their effects on orbiting exoplanets, contributing as it does a large fraction of the ultraviolet flux of low mass stars ($\approx$30--70\,\%, \citealt{franceetal13-1}); but extremely difficult to observe, occulted almost entirely by a combination of absorption by hydrogen in the local interstellar medium (ISM) and geocoronal airglow emission from the Earth's atmosphere. Airglow is very uniform on the angular size scale of stars observed with astronomical ultraviolet spectrometers, and its effects can generally be corrected for by subtracting the off-source background spectrum. However, the ISM is so opaque to \lya\ photons that, near the rest wavelength of the line in the frame of reference of the intervening ISM, even emission from the nearest stars is completely absorbed. Nevertheless, observations of the \lya\ line are the focus of a large number of papers and observing programs, returning data from hundreds of stars to date. These observations have been used to, for example, measure stellar winds \citep{woodetal21-1}, estimate the extreme ultraviolet flux of stars \citep{linskyetal14-1}, characterise the outer heliosphere \citep{woodetal14-1}, and model the atmospheres of the TRAPPIST-1 planets \citep{bourrieretal17-2, wunderlichetal20-1}. 

Stellar \lya\ observations rely on the fact that the \lya\ lines of their targets are broad enough that the wings of the line are detectable on one or both sides of the region of strong ISM absorption and/or airglow. The intrinsic profile and flux of the \lya\ line can then be reconstructed from the wings. Multiple methods have been developed for the reconstruction, such as using metal lines to characterise and thus remove the ISM \citep{woodetal05-1} and/or fitting model line profiles to the wings \citep{bourrieretal15-1,youngbloodetal16-1}, with broad agreement between techniques and recipes. However, statistical uncertainties in the reconstructions, which range from 5\,\% to 100\,\% depending on the data quality and whether the unsaturated deuterium absorption line from the ISM is spectrally resolved, are dominated by degeneracies between the ISM absorbers and intrinsic stellar profile, as well as our incomplete knowledge of the intrinsic \lya\ profile shape for the vast majority of stars. Improving the accuracy of \lya\ reconstructions is particularly important for accurately modeling chemistry in exoplanet atmospheres. Small changes of $\sim$20\,\% in the reconstructed \lya\ flux can propagate to $\sim$30\,\% changes in the O$_2$ and O$_3$ column depths in Earth-like planets orbiting M\,dwarfs \citep{seguraetal07-1}. In mini-Neptune atmospheres, \lya\ is the dominant driver of photochemistry in the atmospheric layers most likely to be probed by future direct observations \citep{migueletal15-1}.

Testing the absolute accuracy of the \lya\ reconstruction routines is challenging, as the ground truth of occulted \lya\ profiles cannot be obtained. In rare cases the \lya\ line can be fully observed for stars with sufficiently high radial velocities to shift the line out of the airglow and the deepest ISM absorption, the best example being Kapteyn’s star at 245\,km\,s$^{-1}$ \citep{guinanetal16-1, schneideretal19-1, youngbloodetal22-1, peacocketal22-1}. Unfortunately, high-velocity stars cannot also be observed with the line occulted. What is required is a star with a radial velocity that changes by many 10s of km\,s$^{-1}$ over time, allowing observations to be taken at low velocities when the \lya\ line is occulted, and the reconstruction based on those data compared with high-velocity, unobscured \lya\ observations of the same star. Such conditions exist in detached Post Common Envelope Binaries (PCEBs), specifically those containing a white dwarf with a main-sequence companion. 

\eguma\ is a PCEB comprising an M4-type dwarf star and a white dwarf in a 16 hour orbit\citep{lanning82-1, bleachetal00-1} at a distance of $\approx 28$\,pc. The white dwarf is unusually cool among the sample of well-studied PCEBs, with an effective temperature of only around 13000\,K \citep{sionetal84-1}, such that the ultraviolet emission from the system is not completely dominated by the white dwarf and shows a detectable contribution from the M\,dwarf. The short orbital period provides the large range of radial velocities (up to $\approx150$\,\kms) that allows the system to be observed at multiple levels of \lya\ occultation, providing a comprehensive test of \lya\ reconstruction routines. This paper presents the results of this experiment using phase-resolved Hubble Space Telescope (HST) spectroscopy of \eguma.

\section{Observations} \label{sec:obs}

\begin{table*}
\centering
\caption{Summary of observations. Dataset numbers are given for retrieval from MAST (\url{https://archive.stsci.edu/hst/}).}  
\begin{tabular}{lcccccccc}
\hline
% & & & Central &  & & & \\
% Date & Instrument & Grating & Wavelength (\AA) & Start Time (UT) & Exposure Time (s) & Dataset & Orbital Phase \\

Date & Instrument & Grating & Central & Start Time (UT) & Exposure Time (s) & Dataset & Orbital Phase \\
& & & Wavelength (\AA) &  & & & \\
\hline 
2017-12-05 & COS & G130M & 1291 & 17:30:29 & 1969 & LDLC05010 & -- \\
2021-04-02  &	STIS &	G140M &	1222 & 06:39:15 & 1358 & OEHUA4010 & 0.53 \\
2021-04-02  &	STIS &	G140M & 1222 & 09:51:27 & 1358 & OEHUA1010 & 0.73 \\
2021-04-02  &	STIS &	G140M & 1222 & 14:35:35 & 1358 & OEHUA2010 & 0.03 \\
2021-04-26  &	STIS &	G140M & 1222 & 19:46:59 & 1358 & OEHUB3010 & 0.3\\
\hline
\hline
\end{tabular}
\label{tab:hst_obs}
\end{table*}

\subsection{HST/COS} \label{sec:cos}
Table \ref{tab:hst_obs} provides a log of the observations of \eguma\ discussed in this section. All data used in this paper can be found in MAST: \dataset[10.17909/k8h7-s556]{http://dx.doi.org/10.17909/k8h7-s556}. \eguma\ was observed with the Cosmic Origins Spectrograph \citep[COS,][]{greenetal12-1} onboard HST as part of program ID 15189. A single observation was obtained on 2017~December~05 with an exposure time of 1969\,s, using the Primary Science Aperture and the G130M grating with a central wavelength of 1291\,\AA. The spectrum was automatically extracted using the standard \texttt{CalCOS} tools. The spectrum is shown in Figure \ref{fig:cos}.
\newpage
\subsection{HST/STIS} \label{sec:stis}

Four additional HST spectra of \eguma\ were obtained using the Space Telescope Imaging Spectrograph \citep[STIS,][]{woodgateetal98-1} on 2021~April~02 (three spectra) and 2021~April~26 (one spectrum)\footnote{This observation was initially attempted on 2021~March~30, but a technical issue caused a complete data loss. We note it here to avoid confusion for readers wishing to retrieve the data. The failed observation is archived on MAST as dataset OEHUA3010.} under program ID 16449. Each spectrum had an exposure time of 1358\,s and was obtained using the G140M grating with a central wavelength of 1222\,\AA\ and the 52X0.1 arcsecond slit. The spectra were timed to obtain data at the 0.0, 0.25, 0.5, and 0.75 phases of the binary orbit based on the ephemeris of \citet{bleachetal02-2}, with an allowable phase error of $\pm 0.05$. The observations were further scheduled such that the radial velocity of the Earth relative to the predicted ISM velocity ($1.45\pm 1.37$\,\kms, \citealt{redfield+linsky08-1}) along the line of sight to \eguma\ was no more than $\pm 20$\,km\,s$^{-1}$, placing the bright geocoronal \lya\ airglow line in the deepest region of ISM absorption. The airglow contribution was further reduced by the use of the narrow 0.1\,arcsecond slit. The spectral trace was automatically located and extracted by the standard {\tt CalSTIS} pipeline. An example STIS spectrum is shown in the top panel of Figure \ref{fig:stis_lya}. We used the {\sc stistools splittag} routine to extract time-series spectra from each observation, and found no evidence for variability of the \lya\ line during any of the visits. 

We evaluated the flux calibration of the STIS spectra by comparison with the COS spectrum. Three of the four spectra showed good agreement at all wavelengths, apart from the expected Doppler shift. Spectrum OEHUA2010 showed significant, unphysical departures from the COS spectrum at wavelengths $<1210$\,\AA, including a region of negative flux and a region roughly 1.5 times higher than the COS flux. We attribute the poor extraction to ``FUV glow'', an irregular region of dark current on the STIS MAMA detector that grows over time after the detector is powered on after an SAA passage\footnote{\url{https://hst-docs.stsci.edu/stisihb/chapter-7-feasibility-and-detector-performance/7-5-mama-operation-and-feasibility-considerations\#id-7.5MAMAOperationandFeasibilityConsiderations-7.5.2MAMADarks}}. OEHUA2010 was the third of three \eguma\ observations on the same day, and the irregular dark current had increased such that the standard background regions ($\pm 300$ pixels from the spectral trace) were no longer representative of the background around the spectral trace. We re-extracted OEHUA2010 with the {\sc stistools x1d} routine, changing the background regions to $\pm 50$ pixels from the trace, which brought the spectrum into much better agreement with the COS spectrum, although with a lower S/N ratio at the shorter wavelengths compared to the other STIS spectra.   
% \pagebreak
\section{COS spectral analysis and experiment design}\label{sec:spectrum}
\begin{figure}
    \centering
    \includegraphics[width=\columnwidth]{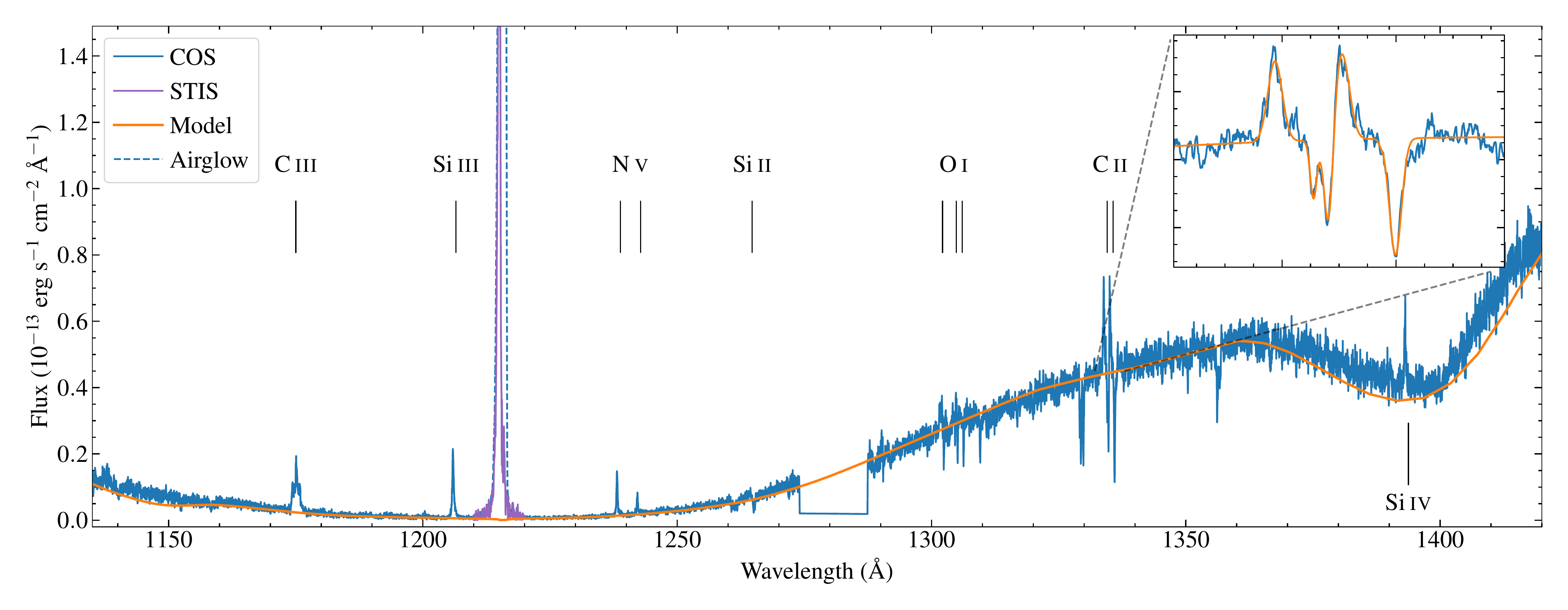}
    \caption{Far ultraviolet spectrum of \eguma. The COS spectrum is shown in blue, with a dashed line indicating the region affected by geocoronal \lya\ airglow. Prominent emission (from the M dwarf atmosphere) and absorption (stellar wind accretion onto the white dwarf) features are labelled. The white dwarf model spectrum is shown in orange. The STIS spectrum shows the highest-velocity and mostly unobscured \lya\ line at Phase 0.73, which exceeds the axis limit by factor $\approx2$. The inset shows the region around the \ion{C}{2} 1335\,\AA\ lines, simultaneously fit with an {\sc astropy} model to measure the instantaneous radial velocity of the M\,dwarf (emission features), ISM (weak absorption features) and the white dwarf (strong absorption features).}
    \label{fig:cos}
\end{figure}

%do we need to explain in full the COS fitting? Don't think so. 
Figure \ref{fig:cos} shows the far-ultraviolet spectrum of \eguma. The spectrum is dominated by the extremely broad \lya\ absorption feature in the white dwarf photosphere, along with the satellite feature around 1400\,\AA\ \citep{allardetal94-1, koesteretal14-1}. Although not strictly correct, we refer to this as the white dwarf continuum hereafter to avoid confusion with the M\,dwarf \lya\ emission line. The spectrum was fit with synthetic models generated by the latest version of the code described in \citet{koester10-1}, using the parallax measurements from Gaia DR2 \citep{gaia18-1} to constrain the distance. As the results are of marginal relevance to this work we do not detail the fitting process here, but point the reader to \citet{wilsonetal21-2} for a full description of the methods used to fit the data from program 15189. We find an effective temperature \Teff$=12599\pm13$\,K and \logg$=7.83\pm0.01$ (statistical uncertainties  only) and infer a white dwarf mass of $\approx0.51$\Msun. Our \Teff\ is substantially lower than the $\approx 18000$\,K found in recent studies \citep[e.g.,][]{gianninasetal11-1, limogesetal15-1}, but those results were based on optical data that feature significant flux contributions from the companion, which can seriously affect the standard analysis of the Balmer absorption lines of the white dwarf. \citet{sionetal84-1} found \Teff$=13000\pm500$\,K based on data from the International Ultraviolet Explorer, consistent with our result.

\defcitealias{astropy18-1}{Astropy Collaboration, 2018}

In addition to the continuum, multiple emission and absorption features are detected. The emission features originate from the outer atmosphere of the M\,dwarf, with the same collection of lines seen in ultraviolet spectra of isolated mid-M\,dwarfs \citep[see for e.g.,][]{loydetal16-1}. The absorption features originate in the white dwarf photosphere and are produced by accretion of the M\,dwarf stellar wind onto the white dwarf \citep{debes06-1}. Of particular interest is the \ion{C}{2}\,doublet around 1335\,\AA\ which shows both emission and absorption features, as well as weak ISM absorption (inset in Figure \ref{fig:cos}). We fit the \ion{C}{2} lines using the {\sc astropy} \citepalias{astropy18-1} modelling functions, using a 2nd order polynomial for the continuum and six Gaussian profiles for the lines, with the line separation and widths fixed between each pair of features. Although only one ISM feature is reliably detected, we find a line-of-sight ISM velocity of $3.7\pm2.8$\,\kms, consistent with the $1.45\pm 1.37$\,\kms\ velocity predicted by the ISM maps from \citet{redfield+linsky08-1} and significantly less than the COS wavelength solution accuracy of $\pm7.5$\,\kms. The \ion{O}{1} lines at 1300\,\AA\ also show the same emission-ISM-absorption configuration, but as these lines are blended with nearby Si lines and geocoronal \ion{O}{1} airglow we did not attempt to fit them. 

The combination of strong detected emission lines and high radial velocity amplitude suggests that this is an ideal system to test the \lya\ reconstruction routines by observing the \lya\ line at different velocities and therefore different levels of obscuration by ISM absorption. Unfortunately the large aperture of COS means that the stellar \lya\ line is completely swamped by bright geocoronal emission over the full radial velocity range (dashed line in Figure \ref{fig:cos}). The experiment was therefore performed using STIS with a narrow (0.1 arcsecond) slit that greatly reduced the airglow contribution.

\section{STIS results} \label{sec:stis_results}

\begin{figure}
    \centering
    \includegraphics[width=\columnwidth]{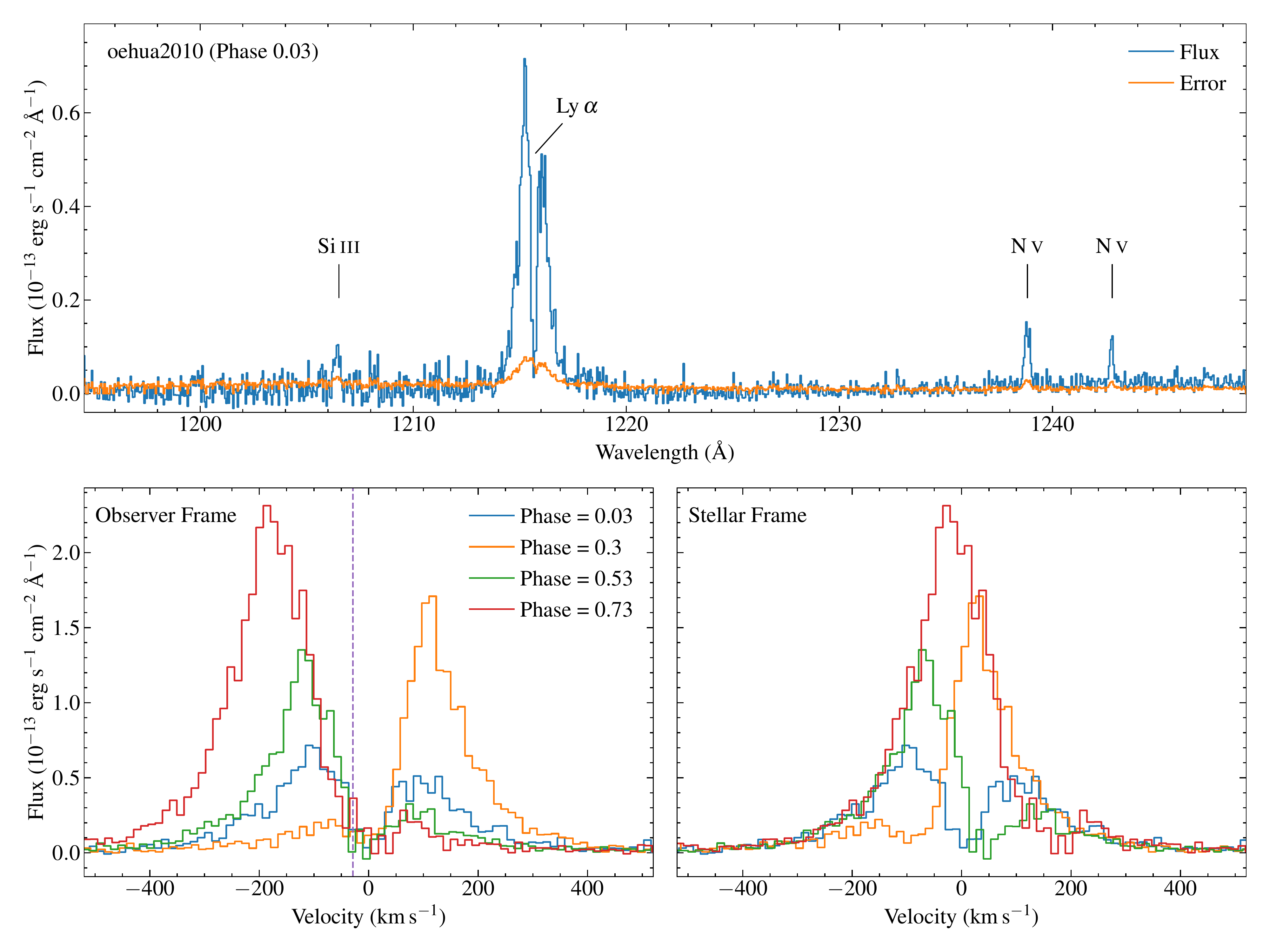}
    \caption{Top panel: Example STIS G140M spectrum of \eguma\ (dataset OEHUA2010) with major emission features marked. Note the increased scatter noise at wavelengths $\lesssim 1214$\,\AA\ due to FUV glow. Bottom panels: The \lya\ emission line in all four spectra, in the observer rest frame (left) and velocity-shifted to the M\,dwarf rest frame (right). The dashed purple line in the left panel shows the center-of-mass radial velocity of the binary system.}
    \label{fig:stis_lya}
\end{figure}

As expected, the phase-resolved STIS observations returned four distinctly different \lya\ profiles. Figure \ref{fig:stis_lya} shows the region around the \lya\ line, velocity adjusted to the observer frame (left) and M\,dwarf rest frame (right). The observations clearly demonstrate the effects of ISM absorption on the \lya\ line, with the line being much fainter at low velocities than higher. The STIS G140M spectra also cover the \ion{Si}{3}\,1206\,\AA\ and \ion{N}{5}\,1240\,\AA\ emission lines, which are clearly detected in all four spectra. 

Given the allowed uncertainty in the phase targeting and the cycle-count errors induced by the twenty-year gap between these observations and those of \citet{bleachetal02-2}, the phases of each observation are not quite at the 0, 0.25, 0.5 and 0.75 positions. Our recalculation of the correct phase positions is described in Section \ref{sec:lyaresults}, but to avoid confusion we use the corrected phases (0.03, 0.3, 0.53 and 0.73) in figures and text throughout the paper. \\

\section{Lyman Alpha reconstructions} \label{sec:lya}

\subsection{Line model and reconstruction routine}

The observed \lya\, spectrum is a combination of four components: Geocoronal airglow from the Earth; Absorption from the ISM; Absorption from the white dwarf photosphere; and finally the feature that we are aiming to characterise, emission from the chromosphere of the M\,dwarf. Our observations were timed such that the radial velocity of the Earth coincided with the expected radial velocity of the ISM in the direction of \eguma, ensuring that the airglow emission fell in the deepest region of ISM absorption and would not overlap with the measurable signal from \eguma. Combined with the standard pipeline background subtraction, we find that airglow makes a negligible contribution to the spectra and can be ignored. The flux from the white dwarf in the region around the \lya\ core is much smaller than that from even the wings of the M\,dwarf emission line and can also be neglected. This leaves the ISM absorption and intrinsic stellar emission to be fitted. 

We reconstructed the \lya\ line for each spectrum with the latest version of the publicly-available {\sc lyapy}\footnote{{\sc lyapy} is actively developed at \url{https://github.com/allisony/lyapy}. The version used in this paper is avaiable at \url{https://doi.org/10.5281/zenodo.6949067}.} routines \citep{youngbloodetal16-1, youngbloodetal21-1}. The intrinsic emission from the M\,dwarf is treated as a Voigt profile \citep{mclean94-1} modelled using the {\sc Astropy Voigt-1D} function. \citet{youngbloodetal22-1} demonstrated that significant self-reversal is present at the cores of \lya\ lines observed for high-velocity M\,dwarfs, so we optionally include a second, absorbing Voigt profile at the line core. The broad ($>$100 \kms) ISM contribution is a combination of absorption from \ion{H}{1} and \ion{D}{1}, with rest wavelengths of 1215.67\,\AA\ and 1215.34\,\AA\ respectively, which we also treat as a pair of Voigt profiles. The STIS G140M grating has insufficient resolution to resolve the \ion{D}{1} line, so the ratio of the column density of the absorption features is fixed at $D/H = 1.5\times10^{-5}$ \citep{linskyetal06-1} and the Doppler parameter $b$, which determines the width of the features, fixed as $b_{HI} = 11.5$\,\kms, $b_{DI} = b_{HI}/\sqrt{2}$ \citep{woodetal04-1, redfield+linsky04-1}. The observed spectra can then be modelled by multiplying the normalised ISM profile with the \lya\ line and convolving the resulting spectrum with the STIS G130M line spread function\footnote{\url{https://www.stsci.edu/hst/instrumentation/stis/performance/spectral-resolution}}. The parameters to be fit are therefore the velocity, amplitude and Gaussian and Lorentzian widths of the \lya\ emission line $V_{Ly\alpha}$, $A$, $FWHM_G$ and $FWHM_L$, a dimensionless self-reversal parameter $p$ (if included), along with the hydrogen column density N(\ion{H}{1}) and ISM velocity $V_{\mathrm{ISM}}$.  The intrinsic \lya\ flux ($F_{Ly\alpha}$), which is the key result for most astrophysical applications, is calculated by numerical integration over the Voigt profile. Assuming that the intrinsic \lya\ emission profile has not changed between observations, fits to all four spectra should return the same result for all parameters except for $V_{Ly\alpha}$. $V_{Ly\alpha}$ should in turn be consistent with the velocities measured from the \ion{Si}{3}\,1206.5\,\AA\ and \ion{N}{5}\,1238.8, 1242.8\,\AA\ emission lines in the same spectrum. These lines were also fit with Voigt profiles to measure their fluxes and radial velocities. For the \ion{N}{5} doublet we fixed the line separation, and fixed the ratio of the line strength to two, the ratio of their respective oscillator strengths.       

As the velocity of the M\,dwarf changes over each exposure, the emission lines may be smeared out the changing Doppler shift. We used the phase positions to calculate the change in velocity over each exposure. The change in velocity is $\leq 18$\,\kms\ for all spectra, smaller than the instrumental resolution of $\approx27$\,\kms. Nevertheless, we added a Doppler broadening effect to our fitting routines for our initial \lya\ reconstructions. The differences between reconstructions with and without broadening included were negligible. As including broadening is computationally expensive, we did not include it in further fits. The line is also subject to rotational broadening of $\approx30$\,\kms\ \citep{bleachetal02-2}, but as it is constant with phase we did not include it as a specific parameter in the mode. 

We fit the profiles using a Markov-Chain Monte Carlo (MCMC) method as implemented by {\sc emcee} \citep{foreman-mackeyetal13-1}. {\sc emcee} maximizes the sum of the logarithm of our parameters' prior probabilities and the logarithm of a likelihood function that measures the goodness of fit of the model to the data. We assume uniform priors for all parameters \citep{youngbloodetal16-1} and a Gaussian likelihood function. $\log$ N(\ion{H}{1}) was forced to be $>17$, and if self-reversal was included then $p$ was forced to be $\geq1$. We used 100 walkers, ran for 50 autocorrelation times, and removed a burn-in period.

In addition to fitting each spectrum individually, we also performed a simultaneous fit to all four spectra. In this case the intrinsic \lya\ and ISM profiles were forced to be the same for all four spectra, with only $V_{Ly\alpha}$ varying between reconstructions. Whilst fitting to the \lya\ line at different velocities is impossible for single stars, we can use this joint fit to inform our discussion of the individual fits, as well as take advantage of this unique dataset to get the best possible reconstruction of the \eguma\ \lya\ line.  

\subsection{Reconstruction Results}
\label{sec:lyaresults}

\begin{figure}
    \centering
    \includegraphics[width=\columnwidth]{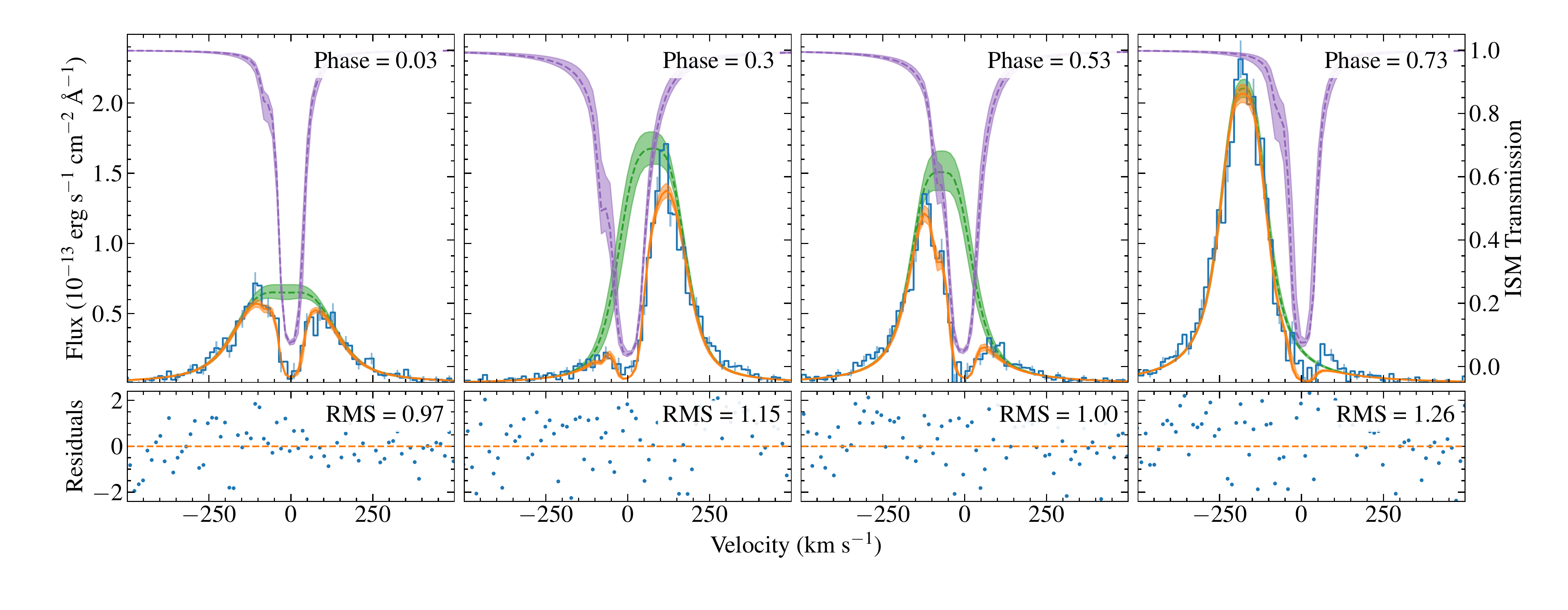}
    \caption{Fits to the \lya\ emission line in all four G140M spectra, ordered by orbital phase. The observed spectrum is shown in blue with errorbars on every third point. The intrinsic \lya\ profile is shown in green and the ISM transmission in purple, both of which are convolved to the instrumental resolution. The product of these profiles is the final fit to the data (orange). Shaded areas behind each profile show the 1\,$\sigma$ uncertainties of the model fits. The bottom panel shows the residuals, calculated as (data-model)/data uncertainty.}
    \label{fig:lyfits}
\end{figure}

\begin{figure}
    \centering
    \includegraphics[width=\columnwidth]{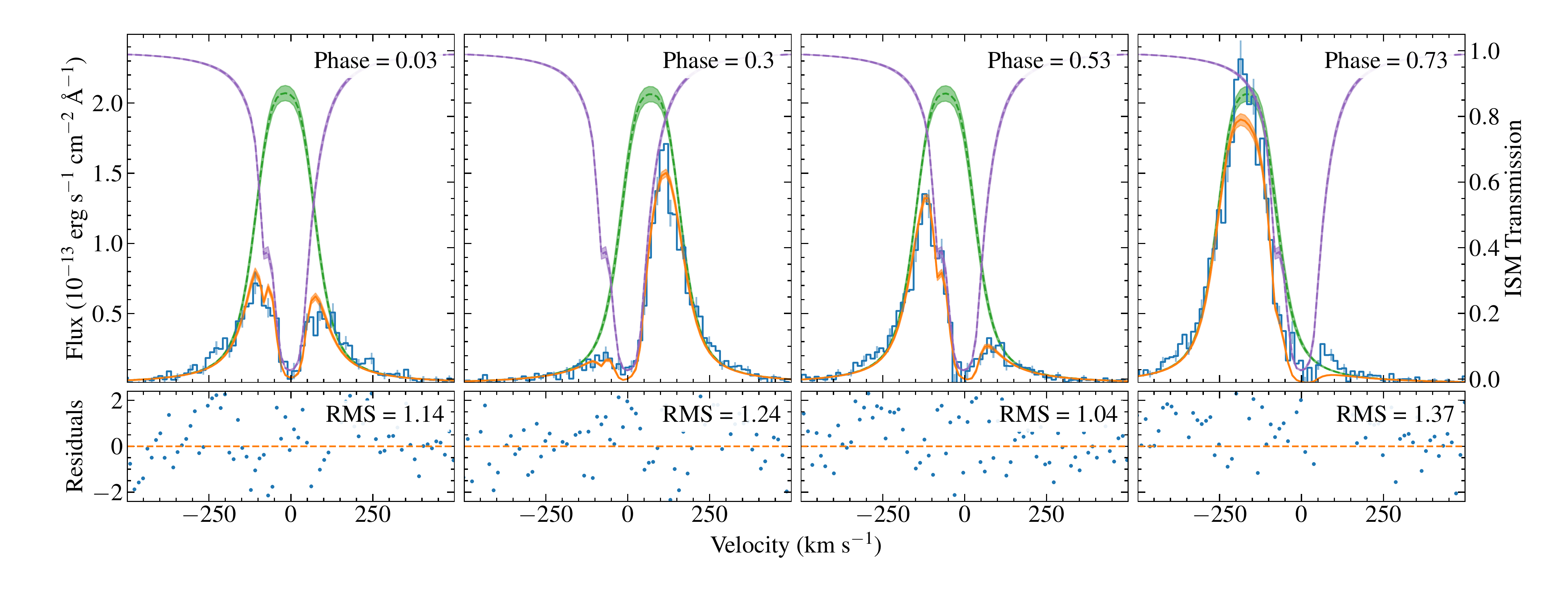}
    \caption{As Figure \ref{fig:lyfits}, but fitting all four spectra simultaneously with only the radial velocity allowed to vary between reconstructions.}
    \label{fig:lyjoint}
\end{figure}

\begin{figure}
    \centering
    \includegraphics[width=\columnwidth]{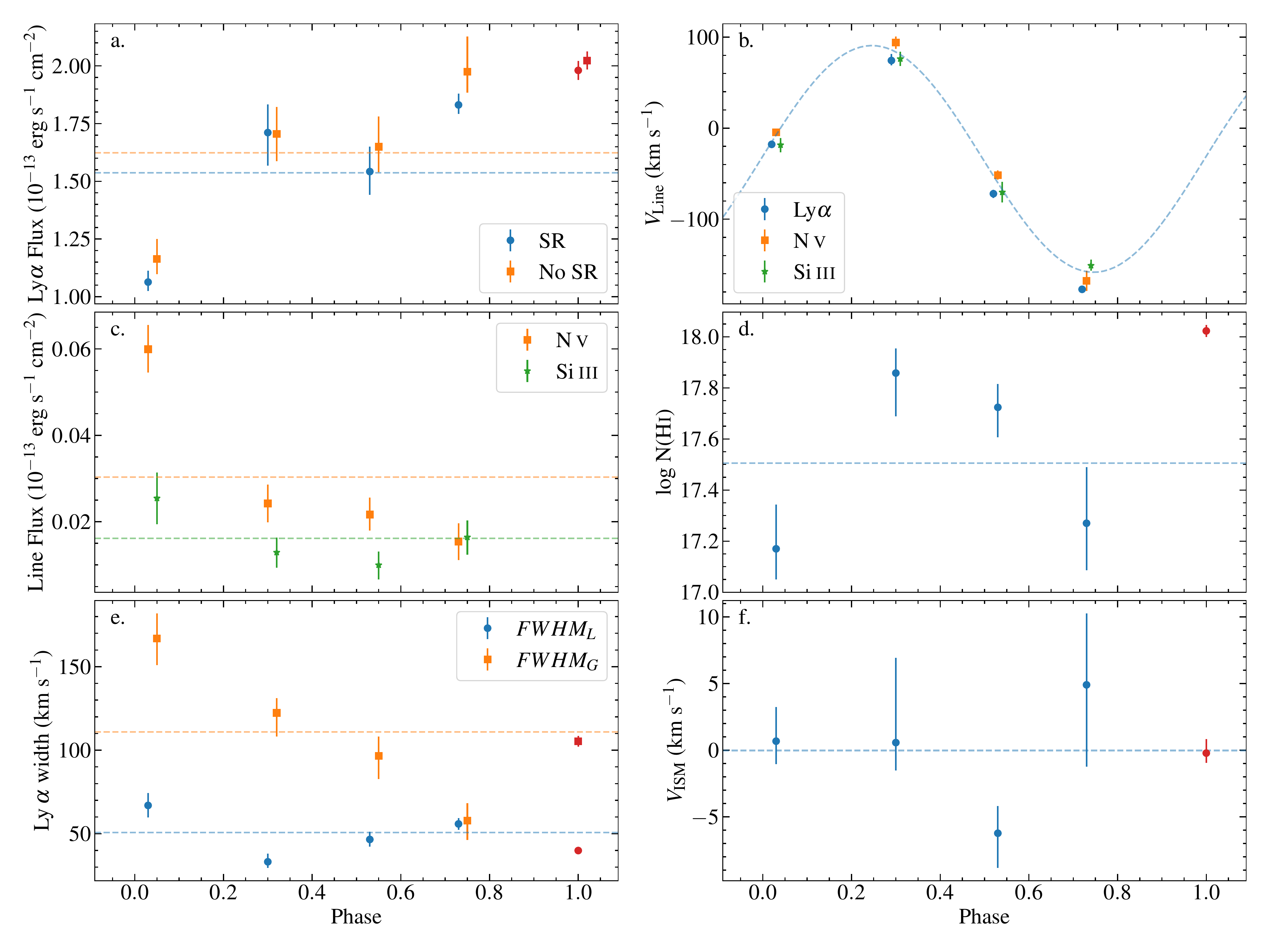}
    \caption{Summary of the reconstruction results: a. The intrinsic flux of the best-fit \lya\ line for each spectrum, comparing fits with and without self reversal (SR); b. The radial velocities of the emission lines; c. The integrated fluxes of the \ion{Si}{3} and \ion{N}{5} emission lines, where \ion{N}{5} is the combined flux of the doublet; d. The ISM hydrogen column density; e. The Voigt widths of the best-fit \lya\ line; and f. the radial velocity of the ISM. The dashed lines show the mean value of each quantity, except for panel b. where the radial velocity curve of the M\,dwarf is shown. In all panels the red points at Phase\,$=1.0$ show the results for the simultaneous fit to all four spectra, which are not included in the means. All results are for fits including self reversal except where noted in panel A. Small phase offsets have been applied where necessary to avoid overlapping points.  }
    \label{fig:summary}
\end{figure}

\begin{figure}
    \centering
    \includegraphics[width=16cm]{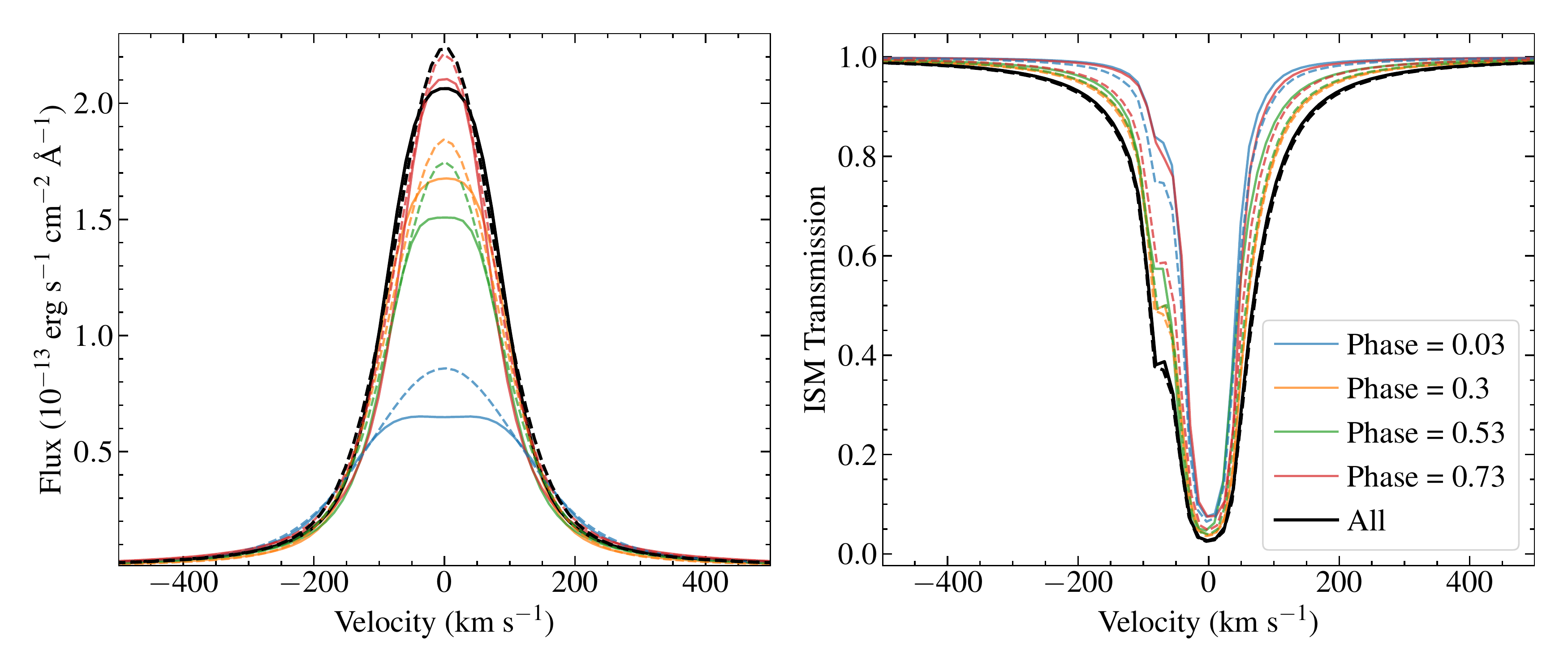}
    \caption{Left: Intrinsic \lya\ line profiles reconstructed from the individual spectra, along with the simultaneous fit to all four, velocity-shifted to the rest frame. Results from fits not including self-reversal are shown as dashed lines. Right: The same for the ISM profile. In both panels the profiles have been convolved with the STIS G140M line spread function. }
    \label{fig:intrinsic_lya}
\end{figure}

The reconstructed \lya\ profiles for each spectrum are show in Figure \ref{fig:lyfits} (individual fits) and Figure \ref{fig:lyjoint} (simultaneous fit). The key parameters are summarised in Figure \ref{fig:summary} and Table \ref{tab:stis_results}. Figure \ref{fig:intrinsic_lya} shows an overlay of the intrinsic \lya\ and ISM profiles.

 For the individual fits, we find that each reconstruction successfully recreates the observed spectrum, with the exception of the possibly spurious feature at $\approx100$\,\kms\ at Phase 0.73. The velocities of the emission lines are also consistent within each spectrum. We fit the radial velocity curve from \citet{bleachetal00-1} to the measured velocities, bounding the orbital velocity amplitude, net velocity and period within their respective uncertainties and allowing the phase to vary freely  (Figure \ref{fig:summary}, panel b). This refit velocity curve was used to calculate the correct phase position of each spectrum, with the corrected phases used throughout this paper. In all cases we find only small differences in fits with and without self reversal, which we discuss in more detail below, but the discussion hereafter will refer to the results from fits including self reversal unless otherwise stated.   

Contrary to our expectations, we find different reconstructed \lya\ emission profiles and intrinsic fluxes for all four spectra, with the peak of the intrinsic profile approximately following the peak of the observed flux. No correlation is seen between the \lya\ flux and the \ion{Si}{3} and \ion{N}{5} fluxes: In fact, the emission line flux is highest when the \lya\ flux is lowest. The reconstructions at Phases 0.3, 0.53 and 0.73 are broadly similar, with intrinsic fluxes within 1--2$\sigma$. However at Phase 0.03, where none of the line core is observed, the intrinsic flux is only $\approx$ 50--70\,\% that found for the other spectra, with a 15\,$\sigma$ difference between the highest and lowest reconstructed fluxes. The $FWHM_G$ is also larger for Phase 0.03. We find that the profile reconstructed from all four spectra simultaneously returns a good fit to all four observed profiles, although the RMS residuals are $\approx5$--20\,\% higher than for the individual fits. Comparing the profiles in Figure \ref{fig:intrinsic_lya}, we find that the intrinsic \lya\ profile is similar to that found when fitting only the least-attenuated observed profile (Phase 0.73) but that a deeper and broader ISM profile is required for a good fit to all four spectra.

The fitted properties of the ISM N(\ion{H}{1}) and $V_{\mathrm{ISM}}$ are within 1--2 $\sigma$ for all four spectra, with the mean $V_{\mathrm{ISM}}$ (1.6\,\kms) in agreement with the $3.7\pm2.8$\,\kms\ measurement from the COS spectrum. However, the spread in the best-fit values of N(\ion{H}{1}) is nearly an order of magnitude, with the lower values likely unphysical (the column density to Alpha Centauri is N(\ion{H}{1}) = 17.6, \citealt{woodetal01-1}). Despite this, swapping the fitted ISM profiles between spectra and calculating the predicted flux returns reasonable matches to the data in all cases. The most likely reason for the wide allowed range of N(\ion{H}{1}) is the low resolution of the G140M grating, and in particular the inability to resolve the \ion{D}{1} line.  

\section{Discussion} \label{sec:disc}

\subsection{Why are the fits different?}
The differences in reconstructed \lya\ flux between the four spectra may have troubling implications for existing \lya\ fits to dozens of stars in the literature. However, we must first rule out that the different results are due to genuine variability and/or consequences of binarity. 

The \eguma\ white dwarf may influence the observations in two ways: Contribution from the (velocity-shifting) underlying spectrum and heating of the facing side of the M\,dwarf. We see no evidence for contributions from the white dwarf spectrum to our fits, as the contribution from the \lya\ emission line wings is over an order of magnitude brighter than the modelled white dwarf spectrum at the same wavelengths, and none of the modelled parameters in Figure \ref{fig:summary} vary in phase with the white dwarf motion (that is, in anti-phase to the M dwarf velocity). The irradiation from the white dwarf is also insufficient to power strong variations in emission line strength. Using equation 7 from \citet{rebassa-mansergasetal13-1} we find that the ratio of the flux generated by irradiation to the inherent flux from the M\,dwarf is $\approx0.5$\,\%, in keeping with the few \% variations seen in optical photometic data \citep[e.g.,][]{bleachetal00-1}. Additionally, if irradiation were causing the flux variations we would expect to see the highest emission line flux at phase 0.53 when the full day side of the M\,dwarf is visible, but the highest fluxes are observed at Phases 0.03 (\ion{N}{5}, \ion{Si}{3}) and 0.73 (\lya). 

All M\,dwarfs demonstrate some level of variability in the ultraviolet \citep{loydetal18-1, franceetal20-1}, including flares that can produce factor 10--100 increases in emission line fluxes for short time periods \citep{froningetal19-1}. The flux of the \ion{N}{5} and \ion{Si}{3} lines are stronger by a factor $\sim2$ at phase 0.03, indicating that \eguma\ may have been undergoing a flare during that exposure. However this does not explain the discrepancy in \lya\ flux, as the response of the \lya\ line to flares is generally small \citep{loydetal18-1}, and the reconstructed \lya\ flux at Phase 0.03 is lower than at other phases, not higher. Explaining the results as genuine variability would also be a strong appeal to coincidence, with the variably exactly mimicking a constant-strength line moving in and out of a region of high ISM attenuation. Finally, the fact that the simultaneous fit \textit{does} return a good fit to all four spectra indicates that there is at least one \lya\ profile that is consistent with all observations. In summary, there is neither firm direct nor circumstantial evidence that the intrinsic \lya\ line profile changed during our observations.

As described in Section \ref{sec:stis}, the spectrum at Phase 0.03 required a custom extraction. To check that our results were not affected by the re-extraction, we performed an additional \lya\ reconstruction based on the original automatic extraction. We found only an $\approx$\,10\,\% change in the reconstructed intrinsic \lya\ flux between the automatic and custom extractions, insufficient to explain the discrepancy with the other observations.

We therefore conclude that the fault is not in our star but in our reconstructions. The differences between the  reconstructions to Phases 0.3, 0.53 and 0.73 are small, so can be discussed together as the high-velocity fits. These fits are clearly incompatible with Phase 0.03. Shifting the intrinsic profile from any of the high-velocity phases to the velocity of Phase 0.03 and attenuating by the ISM from either fit results in a model spectrum that over predicts the observed flux by a factor $\approx2$. Conversely, the intrinsic flux reconstructed from the Phase 0.03 data under predicts the observed flux at the higher velocity observations. 

However, the success of the simultaneous fit shows that there is at least one reconstruction that fits all four phases, and that {\sc lyapy} is able to find it when sufficiently constrained.  Comparing the individual fits to the simultaneous fit suggests an explanation for the different reconstructions: Degeneracy between the strength of the line core and the ISM absorption. The larger intrinsic flux in the simultaneous fit than in the individual fit to Phase 0.03 is offset by a wider ISM profile. With none of the line core observed and the \ion{D}{1} line unresolved, there is no information to distinguish between a weak \lya\ core with low ISM attenuation, or strong \lya\ with proportionally strong attenuation. In the high-velocity fits more of the line core is detected, better constraining the balance between the intrinsic profile and ISM with information that is unavailable at low velocities.

\subsection{Implications for observations at single stars.}
The majority of \lya\ reconstructions are carried out single stars (or individual members of wide binaries) with low radial velocities, based on STIS G140M/L spectra. The spectra of their \lya\ lines appear similar to the observation at Phase 0.03 presented here, with the emission line core completely occulted by the ISM \citep[see for e.g.][]{franceetal13-1, youngbloodetal16-1, bourrieretal18-1, youngbloodetal21-1, wilsonetal21-1}. The published uncertainties for these reconstructions generally fall in the range 5--30\,\%, so the factor two difference found here represents a significant increase in uncertainty. Inaccurate \lya\ measurements propagate into, for example, incorrect integrated far-ultraviolet fluxes and/or inaccurate estimates of the ultraviolet input into the atmospheres of orbiting exoplanets. On the other hand, a factor-two change in \lya\ flux is small in comparison to order-of-magnitude level uncertainties in attempts to determine the flux of the far-ultraviolet continuum \citep{tealetal22-1}. 

If the uncertainly in \lya\ flux is found to have a significant effect on the outcomes of exoplanet atmosphere observations, future \lya\ reconstructions of low velocity data may need to incorporate more detailed priors. For example, the attenuation may be estimated from maps of nearby ISM clouds \citep[e.g][]{redfield+linsky08-1}. Current formula for estimating the \lya\ strength from other emission lines \citep{youngbloodetal17-1, melbourneetal20-1} are too imprecise to improve the priors here, but these relationships may improve with future observations. Although it does not appear to have been a factor in this case, the radial velocity of the \lya\ line could also be constrained via (for example) measurements of the nearby \ion{S}{3} and \ion{N}{5} emission lines. Ideally, higher resolution STIS E140M spectra shoud be obtained, which can resolve the  \ion{D}{1} and tightly constrain the ISM profile, but the lower throughput of E140M makes this impractical for many targets. High-resolution observations of the \ion{Mg}{2} H\&K lines around 2800\,\AA\, obtained contemporaneously with \lya\ observations, can provide additional constraints on the ISM line velocity and column depth \citep[see for e.g.][]{carleoetal21-1}.

\subsection{Self reversal}
Self-reversal occurs in optically-thick emission lines where the line center traces hot, high-altitude, low-density gas that departs significantly from Local Thermodynamic Equilibrium. As a result, thermal emission is less efficient in the line core than the wings, producing to a line profile with a reversed core. \citet{youngbloodetal22-1} performed \lya\ reconstructions for a selection of high radial velocity stars, where the self-reversal of the \lya\ core could be detected. Among their results was that including self-reversal was essential to recover the true \lya\ flux, with reconstructions without self reversal overestimating the \lya\ flux by factors of 40--170\,\%. In contrast to that, our reconstructions to \eguma\ find little difference whether self-reversal is included or not, both in the integrated flux (Figure \ref{fig:summary}\,a.) and profile (Figure \ref{fig:intrinsic_lya}). Spatially resolved observations of active regions on the Sun have consistently shown decreased self reversal in emission line profiles \citep[e.g.][]{schmitetal15-1} in comparison with inactive regions, so the lower self-reversal at \eguma\ could be due to it being more active than the stars characterised by \citet{youngbloodetal22-1}. Alternatively, observational effects could be to blame. Although we found that orbital smearing had no effect on the overall reconstructions, the combination of smearing, rotational broadening and the relatively low resolution G140M grating may have complicated the retrieval of any contribution from self-reversal in the line cores. 

\subsection{Potential future targets}
A caveat to our results is that we only perform this experiment for a single system, and similar observations of other PCEBs may produce different outcomes. The ideal criteria for such observations include: a white dwarf with a wide \lya\ line (i.e., relatively cool) so that the M\,dwarf emission line can be reliably isolated; A radial velocity amplitude $\gtrsim 100$\,\kms\ so that the line peak moves through a range of ISM attenuation; Detectable M\,dwarf emission lines so that an estimate of the \lya\ strength can be made before investing valuable HST time; and a strong enough estimated \lya\ flux to be detected in short, single orbit exposures with the STIS G140M grating, or ideally the higher resolution E140M grating.

Extensive STIS E140M data have already been obtained for the white dwarf-K\,dwarf binary V471\,Tau. Unfortunately the white dwarf makes a considerable, variable contribution to the region around the \lya\ core and extensive post-pipeline data processing is required to isolate the emission line \citep[see for e.g.][]{sionetal12-1, woodetal05-1}. We performed an initial analysis of the V471\,Tau spectra and found that the change in the detected \lya\ emission line morphology is small in comparison to \eguma. Further processing the data to the required level is beyond the scope of this paper, although we recommend this experiment as a secondary science goal in any future analysis of the V471\,Tau dataset. Another PCEB, HZ\,9, was also observed with E140M as part of the same program, but none of the four spectra were obtained at low radial velocities. 

\citet{hernandezetal22-1} present a STIS G140L spectrum of the white dwarf-G\,dwarf binary TYC 110-755-1. Despite a velocity amplitude of only $\approx50$\,\kms\, the \lya\ emission line is clearly detected and distinguishable from the $\approx 16600$\,K white dwarf, as are multiple other emission lines. Further STIS spectroscopy of this star at different phases would allow an excellent test of the \lya\ reconstruction routines at a much earlier spectral type to \eguma.  

In addition to \eguma, \citet{bleachetal00-1} characterised its near-twin PG\,1026+002 (UZ Sex), an M4 spectral type star orbiting a 17000\,K white dwarf with a $\approx200$\,\kms\ radial velocity amplitude. COS spectroscopy of PG\,1026+002 (dataset LDLC03010, Wilson et al. in prep.) shows a similarly broad \lya\ absorption line to \eguma\ and, although no ultraviolet emission lines are detected, \ion{Ca}{2} H\&K and H\,$\alpha$ emission lines are present in optical spectra  \citep{napiwotzkietal20-1} allowing an estimate of the \lya\ strength. PG\,1026+002 is therefore an ideal target to repeat this experiment, comparing \eguma\ with a potentially less active example of a similar spectral type.
% and if I'd noticed this before we could have added it to the proposal, oops. 

\section{Conclusion} \label{sec:concs}
By observing the \lya\ emission of the binary star \eguma\ at multiple velocities, we have found that the widely used {\sc lyapy} reconstruction routines return different results as a function of how much of the \lya\ core is detected, with a difference of factor $\approx2$ between the strongest and weakest reconstruction. We find that this is unlikely to be due to intrinsic variation from the star, both as there is no plausible physical mechanism to provide such variation, and as a simultaneous fit to all observations returns a single consistent result. We suggest that the issue is a degeneracy between the strength of the ISM attenuation and \lya\ flux, especially at low velocities where none of the line core is detected. \lya\ reconstructions using medium-resolution spectra of  single, low-velocity stars may therefore need additional prior constraints on the \lya\ and ISM profiles to provide a precise \lya\ flux measurement.

\begin{acknowledgments}
This research is based on observations made with the NASA/ESA Hubble Space Telescope obtained from the Space Telescope Science Institute, which is operated by the Association of Universities for Research in Astronomy, Inc., under NASA contract NAS 5–26555. These observations are associated with programs 15189 and 16449. Support for program 16449 was provided by NASA through a grant from the Space Telescope Science Institute, which is operated by the Association of Universities for Research in Astronomy, Inc., under NASA contract NAS 5-03127.BTG was supported by grant ST/T000406/1 from the Science and
Technology Facilities Council (STFC). This project has received funding from the European Research Council (ERC) under the European Union’s Horizon 2020 research and innovation programme (Grant agreement No. 101020057). We thank the HST operations team for their work ensuring the accurate phasing of the STIS observations, and Daniel Welty at STScI for assistance with the FUV glow issue.   

\end{acknowledgments}


\vspace{5mm}
\facilities{HST(COS,STIS)}


% \defcitealias{astropy18-1}{Astropy Collaboration, 2018}
\software{astropy \citepalias{astropy18-1}, stistools\footnote{\url{https://stistools.readthedocs.io/en/latest/}}, scipy \citep{virtanenetal20-1}, numpy \citep{harrisetal20-1}, matplotlib \citep{hunter07-1}, emcee \citep{foreman-mackeyetal13-1}}



\appendix
\section{Table of results}
Table \ref{tab:stis_results} shows the results of our \lya\ reconstructions for the four STIS spectra. 
\begin{sidewaystable}

% \begin{table*}
    \centering
    \begin{tabular}{lcccccccccccc}
    \hline
    & & \multicolumn{3}{c}{With Self Reversal} & \multicolumn{2}{c}{Without Self Reversal} & &&& \\
    Dataset &  Phase & $F_{Ly\alpha}$ & $V_{Ly\alpha}$ & $p$ & $F_{Ly\alpha}$ & $V_{Ly\alpha}$ & F$_{Si\,II}$ &$V_{Si\,II}$ & $F_{N\,V}$ &$V_{N\,V}$ & N(\ion{H}{1}) (cm$^{-2}$)& $V_{\mathrm{ISM}}$ \\  \hline

OEHUA2010 & 0.03 & {1.06}$^{+0.05}_{-0.04}$ & {-17.8}$^{+3.4}_{-3.4}$ & {1.15}$^{+0.21}_{-0.11}$ & {1.16}$^{+0.09}_{-0.07}$ & {-19.0}$^{+3.3}_{-3.2}$ & {0.025}$^{+0.006}_{-0.006}$ & {-18.6}$^{+7.7}_{-8.1}$ & {0.060}$^{+0.006}_{--0.029}$ & {-4.6}$^{+3.0}_{-2.8}$ & {17.2}$^{+0.2}_{-0.1}$ & {0.7}$^{+2.6}_{-1.7}$ \\ 
OEHUB3010 & 0.3 & {1.71}$^{+0.12}_{-0.14}$ & {74.3}$^{+7.2}_{-5.2}$ & {1.04}$^{+0.07}_{-0.03}$ & {1.71}$^{+0.12}_{-0.12}$ & {78.7}$^{+5.0}_{-4.4}$ & {0.013}$^{+0.003}_{-0.004}$ & {76.1}$^{+7.8}_{-7.9}$ & {0.024}$^{+0.004}_{--0.007}$ & {94.0}$^{+6.6}_{-7.1}$ & {17.9}$^{+0.1}_{-0.2}$ & {0.6}$^{+6.4}_{-2.1}$ \\ 
OEHUA4010 & 0.53 & {1.54}$^{+0.11}_{-0.10}$ & {-72.1}$^{+4.2}_{-4.8}$ & {1.10}$^{+0.19}_{-0.08}$ & {1.65}$^{+0.13}_{-0.11}$ & {-70.8}$^{+4.0}_{-3.9}$ & {0.010}$^{+0.003}_{-0.003}$ & {-70.4}$^{+11.6}_{-11.1}$ & {0.022}$^{+0.004}_{--0.008}$ & {-51.7}$^{+5.3}_{-4.8}$ & {17.7}$^{+0.1}_{-0.1}$ & {-6.2}$^{+2.0}_{-2.6}$ \\ 
OEHUA1010 & 0.73 & {1.83}$^{+0.05}_{-0.04}$ & {-177.1}$^{+2.3}_{-1.9}$ & {1.06}$^{+0.10}_{-0.05}$ & {1.97}$^{+0.15}_{-0.09}$ & {-171.1}$^{+8.2}_{-4.4}$ & {0.016}$^{+0.004}_{-0.004}$ & {-151.0}$^{+6.4}_{-5.7}$ & {0.015}$^{+0.004}_{-0.005}$ & {-167.9}$^{+11.3}_{-11.0}$ & {17.3}$^{+0.2}_{-0.2}$ & {4.9}$^{+5.4}_{-6.2}$ \\ 
All & -- &{1.98}$^{+0.04}_{-0.04}$ & -- &  {1.01}$^{+0.02}_{-0.01}$ & {2.02}$^{+0.04}_{-0.04}$ & -- &  {0.042}$^{+0.003}_{-0.004}$ & -- & {0.175}$^{+0.005}_{-0.005}$ & -- & {18.0}$^{+0.0}_{-0.0}$ & {-0.2}$^{+1.0}_{-0.7}$ \\

    \hline
    \hline
    \end{tabular}
    \caption{Results of the STIS analysis by dataset. All line fluxes have units of 10$^{-13}$ erg s$^{-1}$ cm$^{-2}$ \AA$^{-1}$, velocities are in \kms.}
    \label{tab:stis_results}
% \end{table*}

\end{sidewaystable}
% Some text here
% \begin{figure}
%     \centering
%     \includegraphics[width=\columnwidth]{eg_uma_stis_lines.pdf}
%     \caption{Emission lines in the STIS spectra (blue), overplotted with the model fits used to measure their radial velocities. Green dashed lines show the rest wavelengths of the emission lines.}
%     \label{fig:stis_lines}
% \end{figure}


\bibliography{aabib}{}
\bibliographystyle{aasjournal}



\end{document}